\documentclass[manuscript]{aastex63}
\usepackage{bm}

\received{July 29, 2020}
\revised{November 30, 2020}
\accepted{November 30, 2020}
\submitjournal{ApJ}

\shorttitle{Evolution of the Non-Potential Field in AR 12673}
\shortauthors{Yamasaki et al.}

\begin{document}

\title{Evolution of the Non-potential Magnetic Field in the Solar Active Region 12673 Based on a Nonlinear Force-Free Modeling}

\email{dyamasaki@kusastro.kyoto-u.ac.jp}

\author{Daiki Yamasaki}
\affiliation{Department of Astronomy, Kyoto University \\
Kitashirakawaoiwake-cho, Sakyo-ku, Kyoto, 606-8502, Japan}

\author{Satoshi Inoue}
\affiliation{Institute for Space-Earth Environmental Research (ISEE), Nagoya University \\
Furo-cho, Chikusa-ku, Nagoya, 446-8601, Japan}

\author{Shin'ichi Nagata}
\affiliation{Astronomical Observatory, Kyoto University \\
 Kurabashira, Kamitakara-cho, Takayama, 506-1314, Japan}

\author{Kiyoshi Ichimoto}
\affiliation{Astronomical Observatory, Kyoto University \\
Kurabashira, Kamitakara-cho, Takayama, 506-1314, Japan}



\begin{abstract}
  Active region (AR) 12673 produced many M-class and several X-class flares, one of which being an X9.3 flare, which is recorded as the largest solar flare in solar cycle 24.
  We studied the evolution of the three-dimensional flare-productive magnetic field within AR 12673, using a time series of nonlinear force-free field extrapolations of every 12 hours from September 4th 00:00 UT to 6th 00:00 UT.
  Our analysis found that three magnetic flux ropes (MFRs) are formed by September 4th, one of which produced the X9.3 flare on September 6th.
  One MFR has positive magnetic twist which is a different sign from other two MFRs.
  Since the temporal evolution of the magnetic flux of the MFR accumulating the positive twist is consistent with the profile of the $GOES$ X-ray flux well observed from September 4th to 6th, we suggest that the formation of the MFR having the positive twist is closely related to the occurrence of the M-class flares including an M5.5 flare.
  We further found a magnetic null in the magnetic field surrounding the MFRs, in particular, above the MFR having positive twist before the M5.5 flare which is the largest M-flare observed during this period.
  By comparing with the AIA 1600 $\mathrm{\AA}$ images, we found that the footpoints of the overlying field lines are anchored to the area where the brightening was initially observed.
  Therefore, we suggest that reconnection induced by the torus instability of the positively twisted MFR at the null possibly drived the M5.5 flare.
\end{abstract}



\section{Introduction} \label{sec:intr}
Solar flares are widely considered to be releases of the magnetic energy accumulated in the solar corona.
They are often observed in solar active regions (ARs) where the magnetic fields are highly deviated from the potential magnetic field that corresponds to the lowest energy state \citep{Priest2002}.
Shearing and converging motions of the photosphere highly deform the coronal  magnetic field, resulting in accumulated free magnetic energy. 
The magnetic flux rope (MFR), which is a bundle of helical magnetic field lines, is formed in the lower corona via flux emergence, photospheric motions, and reconnection \citep{vanBallegooijen1989,Okamoto2008,Filippov2015,Cheng2017}.
Since the MFR accumulates the free magnetic energy, it becomes a source of solar eruptions. 
The eruption of the MFR drives magnetic reconnection, which plays an important role in the energy release of the solar flares \citep{Shibata2011}, and coronal mass ejections \citep[CMEs; ][]{Chen2011,Schmieder2015}.\\
~ AR 12673 came at the end of August 2017 and was the most flare-productive AR in solar cycle 24.
It produced many M-class flares and several X-class flares from September 2nd to 10th, among which an X9.3 flare was the largest, and produced a geo-effective CME \citep{Yan2018,Shen2018,Chertok2018,Soni2020}. 
AR 12673, at first, appeared as a monopole sunspot observed at the east solar limb in August 29th.
On September 3rd and 4th, two dipoles emerged beside the pre-exisiting sunspot, and the AR evolved into the complex $\delta$ sunspots \citep{Yang2017}. 
During the sunspot evolution in the AR, the negative polarity strongly intruded into the neighboring positive polarity before the X2.2 flare \citep{Bamba2020}, therefore this intrusion would play an important role in causing the X-flares.
While these studies have been done, \citet{Inoue2018b} performed the data-constrained magnetohydrodynamics simulation focusing on X-class flares observed in September 6th, and proposed that the erupting magnetic flux is formed by reconnection between twisted field lines existing above PIL before the flares. \\
~ As we mentioned above, many studies focus on the evolution of the AR and the onset mechanisms and dynamics of the X2.2 and X9.3 flares in terms of observational and numerical approaches. 
However, the overall evolution of the 3D magnetic field in this AR and flare activities before the ocurrence of the X-class flares are not investigated well. 
In this study, we investigate the evolution of the 3D magnetic field using a nonlinear force-free field (NLFFF) extrapolated from the photospheric magnetic field, and discuss the initiation of M-class flares observed  before September 6th.
We focus on the M5.5 flare observed at 20:30 UT on September 4th.
Since this flare produced a geo-effective CME, our study is important also for the space weather forecaset.
We eventually discuss a relationship between the M-class flare activities and the X-class flares which come later. \\
~ The rest of this paper is structured as follows:
the observations and methods of analysis are described in Section \ref{sec:meth}, results are presented in Section \ref{sec:resu}, and important discussions arising from our findings are summarized in Section \ref{sec:disc}.

\section{Methods} \label{sec:meth}
\subsection{Observation} \label{subsec:obse}
\begin{figure*}[htb]
\begin{center}
\includegraphics[bb= 0 0 837.6 763.68, width=150mm]{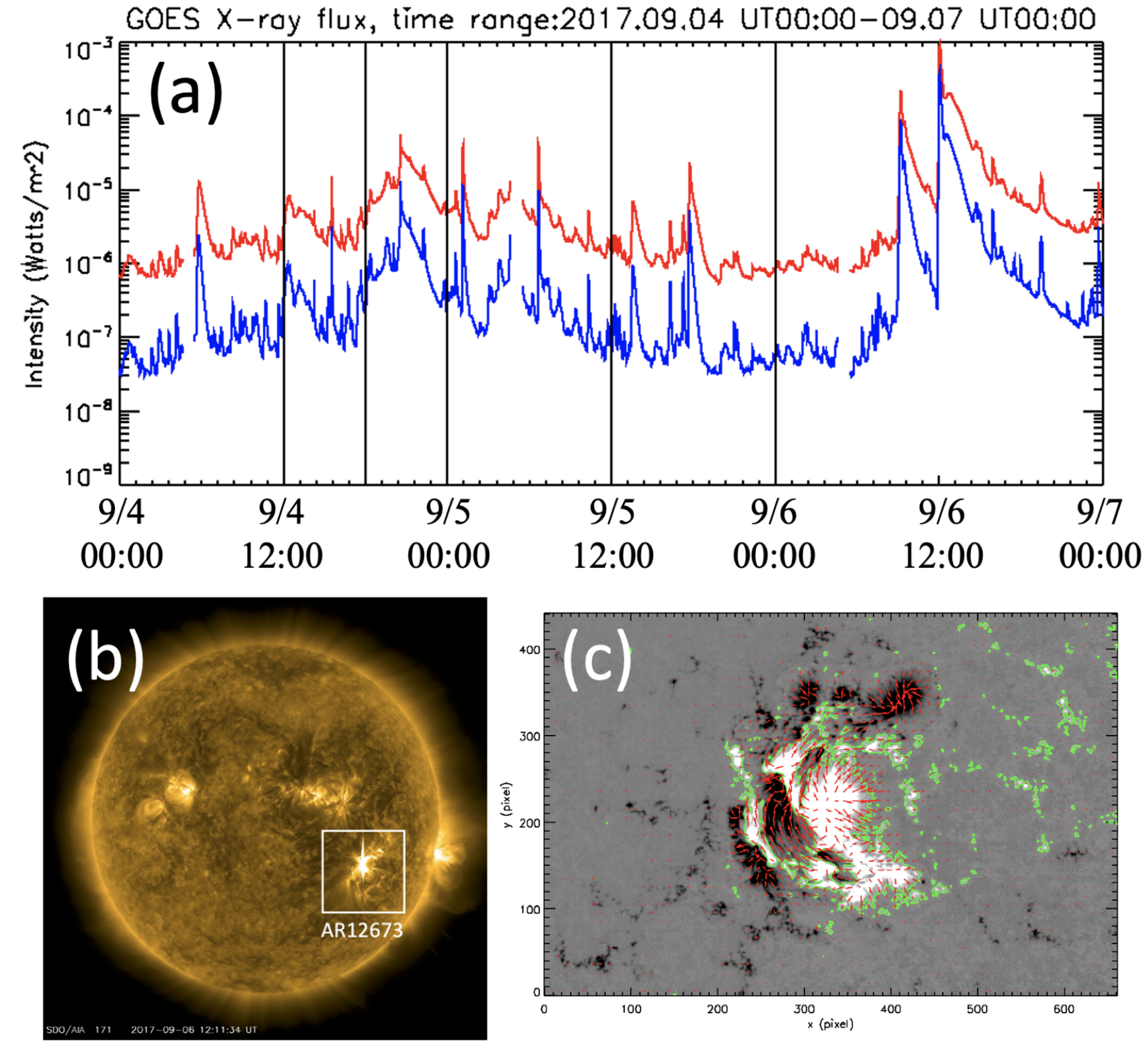}
\caption{(a) Time profile of the soft X-ray flux between 2017 September 4th and 7th measured by the $GOES$ 13 satellite. Red and blue lines correspond to the solar X-ray emission in the $1-8\mathrm{\AA}$ and $0.5-4.0\mathrm{\AA}$ passbands, respectively. Several M-class and X-class flares were observed in this AR. (b) Extreme-UV image of whole Sun observed at 171$\mathrm{\AA}$ at the time of September 6th 12:11 UT when the X9.3 flare was occurring. The white box shows the location of AR 12673. (c) Photospheric vector magnetic field taken at September 4th 18:00 UT. Background gray scale and red arrows show the vertical and the horizontal component of the magnetic field respectively. Green lines show the PIL.}
\label{fig1}
\end{center}
\end{figure*}

As is shown in the $GOES$ X-ray flux profile in figure \ref{fig1} (a), AR 12673 produced 12 M-class flares between 2017 September 4th and 5th, prior to X9.3 on September 6th \citep{Yang2017}.
In figure \ref{fig1} (b), we show the extreme ultraviolet image of the whole Sun on September 6th 12:11 UT in 171$\mathrm{\AA}$ observed by the Atmospheric Imaging Assembly \citep[AIA; ][]{Lemen2012} on board the $Solar~ Dynamics~ Observatory$ \citep[$SDO$; ][]{Pesnell2012}. 
In the figure, AR 12673 is highlighted by the white box.
Figure \ref{fig1} (c) shows the photospheric vector magnetic field of AR 12673, observed at 18:00 UT on September 4th taken by the Helioseismic and Magnetic Imager \citep[HMI; ][]{Scherrer2012} on board the $SDO$.
Details of the vector magnetic field data reduction and other related information about HMI data products can be found in \citet{Hoeksema2014} and \citet{Bobra2014}.
In this image, which was taken approximately 2 hours before the M5.5 flare, 3 polarity inversion lines (PILs) and the sheared magnetic field lines along PILs can be identified.

\subsection{Nonlinear force-free field extrapolation} \label{subsec:nlff}
\begin{figure*}[htb]
\begin{center}
\includegraphics[bb= 0 0 583.69 564, width=160mm]{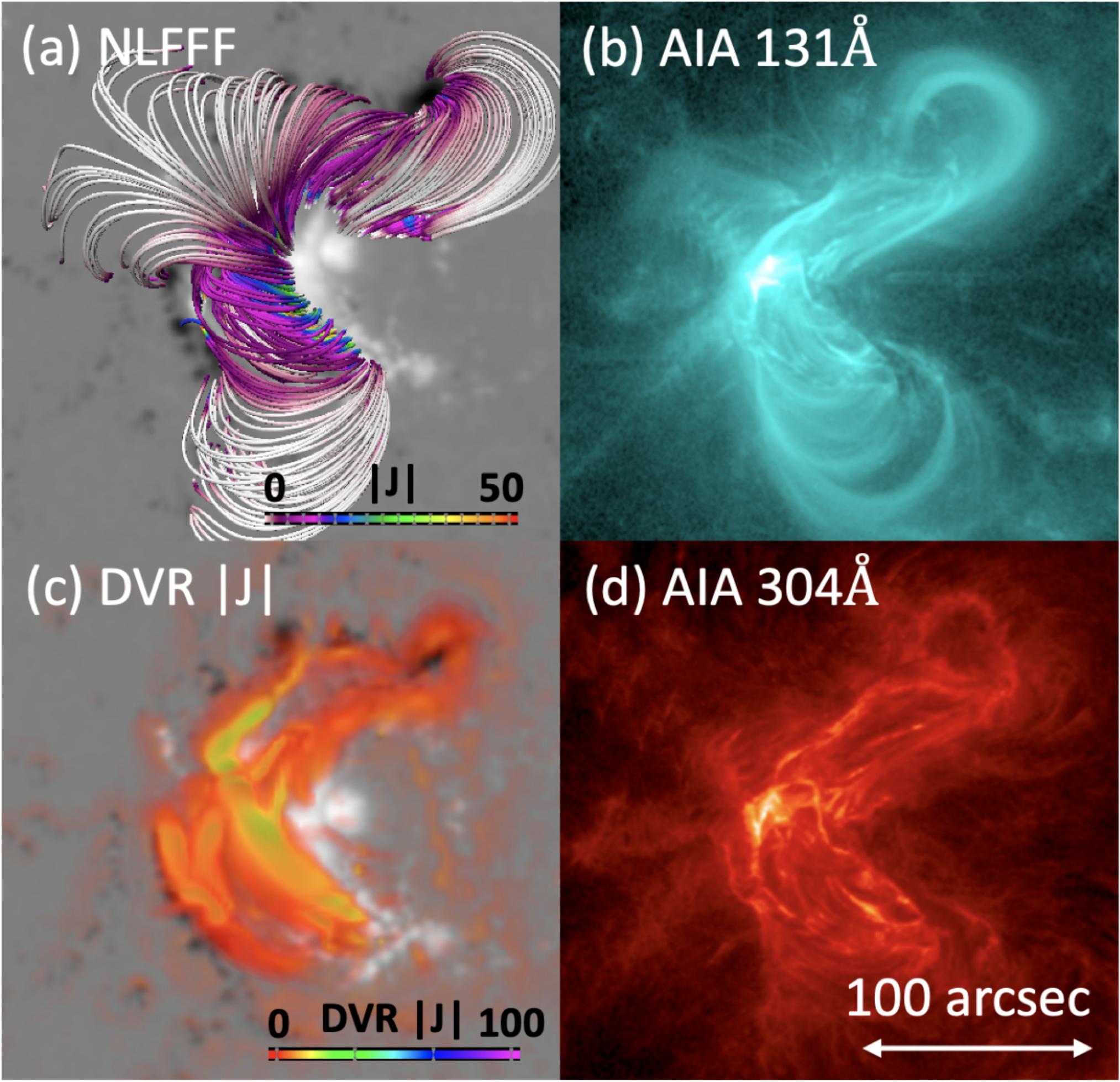}
\caption{(a) Three-dimensional coronal magnetic field strucuture obtained from the NLFFF extrapolation at 18:30 UT on September 4th. This time is approximately 2 hours before the M5.5 flare. (b) AIA $131$ $\mathrm{\AA}$ image taken at 18:30 UT on September 4th. (c) The electric current distribution calculated from the NLFFF extrapolation at 18:30 UT on September 4th. (d) AIA $304$ $\mathrm{\AA}$ image taken at 18:30 UT on September 4th.}
\label{figX}
\end{center}
\end{figure*}

Current observations cannot measure the coronal magnetic field directly.
In order to obtain 3D magnetic fields in the corona, we performed a nonlinear force-free field extrapolation \citep{Wiegelmann2012,Inoue2016,Guo2017}.
NLFFF extrapolations are carried out under the following assumptions: the plasma $\beta$ in the solar corona is small (0.01-0.1), thus the gas pressure and gravity can be neglected to the Lorentz Force.
The time scale of photospheric motions are much longer than that of the Alfven time scale, which characterizes the response time of the coronal magnetic field, so that the coronal magnetic field approximately evolves under a quasi-static state.
Therefore, the evolution of the coronal magnetic field can be approximated by a series of NLFFF.
In this study, we set the observed photospheric vector magnetic fields, which are preprocessed according to \citet{Wiegelmann2006}, as the bottom boundary condition and calculated the 3D coronal magnetic fields to satisfy the force-free state.
The extrapolation is performed using the following procedure: we first calculate the 3D potential field from the vertical component of the observed photospheric magnetic field according to \citet{Sakurai1982}.
Next, we change the horizontal component of the bottom boundary from the potential field to the observed magnetic field, after that the horizontal components perfectly fit to the observed magnetic field.
We continue the calculation of the coronal magnetic field with the fixed boundary condition until the field converges to the state with the minimum deviation from the force-free state.
In this study, to obtain the force-free field, we employ the MHD relaxation method developed by \citet{Inoue2014a} and \citet{Inoue2016}.
We solved the following equations: 
\begin{eqnarray}
  \rho &=& |\bm{B}|, \label{eq1}\\
  \frac{\partial \bm{v}}{\partial t} &=& -(\bm{v}\cdot{\bm{\nabla}})\bm{v}+\frac{1}{\rho}\bm{J}\times\bm{B}+\nu\bm{\nabla}^{2}\bm{v}, \label{eq2}\\
  \frac{\partial \bm{B}}{\partial t} &=& \bm{\nabla}\times(\bm{v}\times\bm{B}-\eta\bm{J})-\bm{\nabla}\phi, \label{eq3}\\
  \bm{J} &=& \bm{\nabla}\times\bm{B},\\
  \frac{\partial \phi}{\partial t}&+&c_{\mathrm{h}}^{2}\bm{\nabla}\cdot\bm{B} = -\frac{c_{\mathrm{h}}^{2}}{c_{\mathrm{p}}^{2}}\phi, \label{eq5}
\end{eqnarray}
where $\rho$, $\bm{B}$, $\bm{v}$, $\bm{J}$, and $\phi$ are plasma pesudo-density, magnetic flux density, velocity, electric current density, and a conventional potential to reduce errors derived from $\bm{\nabla}\cdot\bm{B}$ \citep{Dedner2002}.
The pseudo-density in equation (\ref{eq1}) is assumed to be propotional to $|\bm{B}|$.
The Alfven velocity is uniform for the density model $\rho \propto |\bm{B}|^{2}$. 
In these equations, the length, magnetic field, density, velocity, time, and electric current density are normalized by $L^*=240.0$ $\mathrm{Mm}$, $B^*=3000$ $\mathrm{G}$, $\rho^{*}=|B^{*}|$, $V_{\mathrm{A}}^{*}\equiv B^*/(\mu_{0}\rho^*)^{1/2}$, where $\mu_{0}$ is the magnetic permeability, $\tau_{\mathrm{A}}^{*}\equiv L^{*}/V_{\mathrm{A}}^{*}$, and $J^{*}=B^{*}/\mu_{0}L^{*}$, respectively.
In equation (\ref{eq2}), $\nu$ is a viscosity fixed at $1.0\times10^3$.
The coefficients $c_{\mathrm{h}}^{2}, c_{\mathrm{p}}^{2}$ in equation (\ref{eq5}) are fixed to the constant values $0.04$ and $0.1$.
The resistivity in equation (\ref{eq3}) is given as $\eta=\eta_{0}+\eta_{1}|\bm{J}\times\bm{B}||\bm{v}|^{2}/|\bm{B}|^{2}$, where $\eta_{0}=5.0\times10^{-5}$ and $\eta_{1}=1.0\times10^{-3}$ in non-dimensional units.
As for the choice of these parameters, see \citet{Inoue2016}. 
The second term of the resistivity is introduced to accelerate the relaxation to the force-free state, particularly in regions of weak field.\\
~ We use the photospheric magnetic field obtained at 2017 September 4th 00:00 UT, 4th 12:00 UT, 4th 18:00 UT, 5th 00:00 UT, 5th 12:00 UT, and 6th 00:00 UT.
In figure \ref{fig1} (a), we indicate the time of each NLFFF calculation with vertical black solid lines.
Figure \ref{figX} shows the 3D coronal magnetic field in (a), the current density obtained from the NLFFF extrapolation in (c) and AIA images in 131 $\mathrm{\AA}$ and 304 $\mathrm{\AA}$, respectively in (b) and (d) at 18:30 UT on September 4th.
In the figures, the extrapolated field lines match those inferred from the AIA 131 $\mathrm{\AA}$ image and also the distribution of the current density is in good agreement with the intensity of the AIA 304 $\mathrm{\AA}$ image.
Therefore, we conclude that the NLFFF extrapolation reproduces the observed field lines well. 

\section{Results} \label{sec:resu}
\subsection{Temporal evolution of the NLFFF between September 4th 00:00 UT and September 6th 00:00 UT}
\subsubsection{Temporal evolution of the photospheric and coronal magnetic fields} \label{subsec:teom}
\begin{figure*}[htb]
\begin{center}
\includegraphics[bb= 0 0 698.88 768.96, width=160mm]{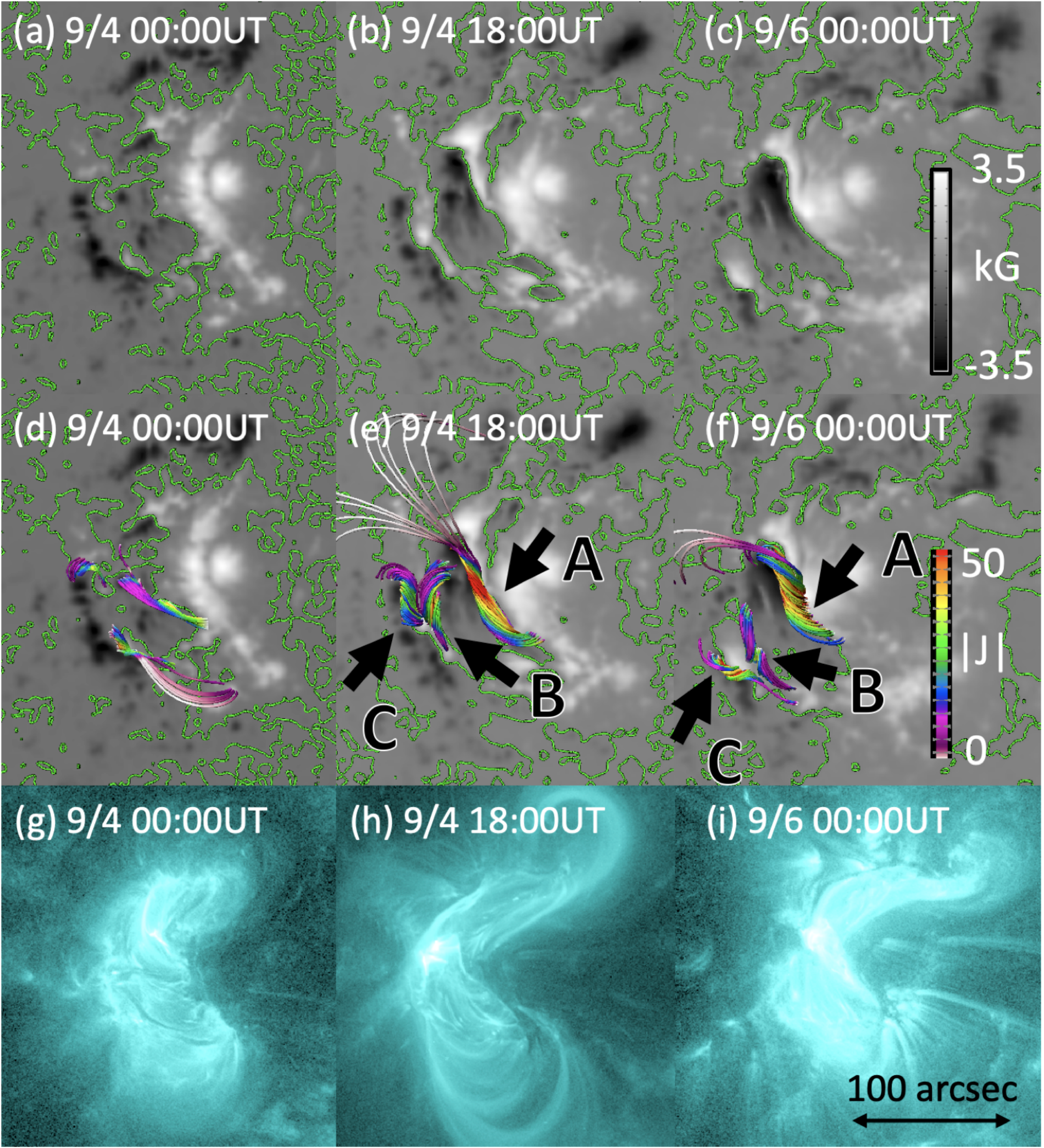}
\caption{Upper panels (a), (b), and (c) show the temporal evolution of the vertical component of the photoshperic magnetic field. Background gray scale and green contour correspond to the vertical magnetic field at the photosphere and PILs, respectively. Middle panels (d), (e), and (f) show the temporal evolution of the MFRs. The color of field lines correspond to the electric current density. Lower panels (g), (h), and (i) shows the EUV $131$ $\mathrm{\AA}$ images observed by the $SDO$/AIA.}
\label{fig2}
\end{center}
\end{figure*}

We first show the temporal evolution of the photospheric magnetic field in figures \ref{fig2}(a)-(c) and the MFRs obtained from the NLFFF extrapolations are shown in figures \ref{fig2}(d)-(f).
At September 4th 00:00 UT, the highly twisted field lines come up according to the photospheric motion.
Furthermore, at 18:00 UT which is approximately 150 minutes before the M5.5 flare peak time, we found an MFR along each of the PILs.
In order to identify these three MFRs, we name MFR A, B, and C as indicated in figure \ref{fig2} (e) and use them in the rest of this paper.
MFR A, which exists at the PIL on the west side of the AR, where the X-class flares were observed on September 6th \citep{Liu2019}, was already formed at 18:00 UT September 4th.
Although the MFRs B and C still exist on September 6th, they look to be relaxed compared to those in 4th.
Therefore, we suggest that these MFRs are involved in causing the M-class flares.

\subsubsection{Temporal evolution of the magnetic twist} \label{subsec:teot}
\begin{figure*}[htb]
\begin{center}
\includegraphics[bb= 0 0 1070.88 705.6, width=160mm]{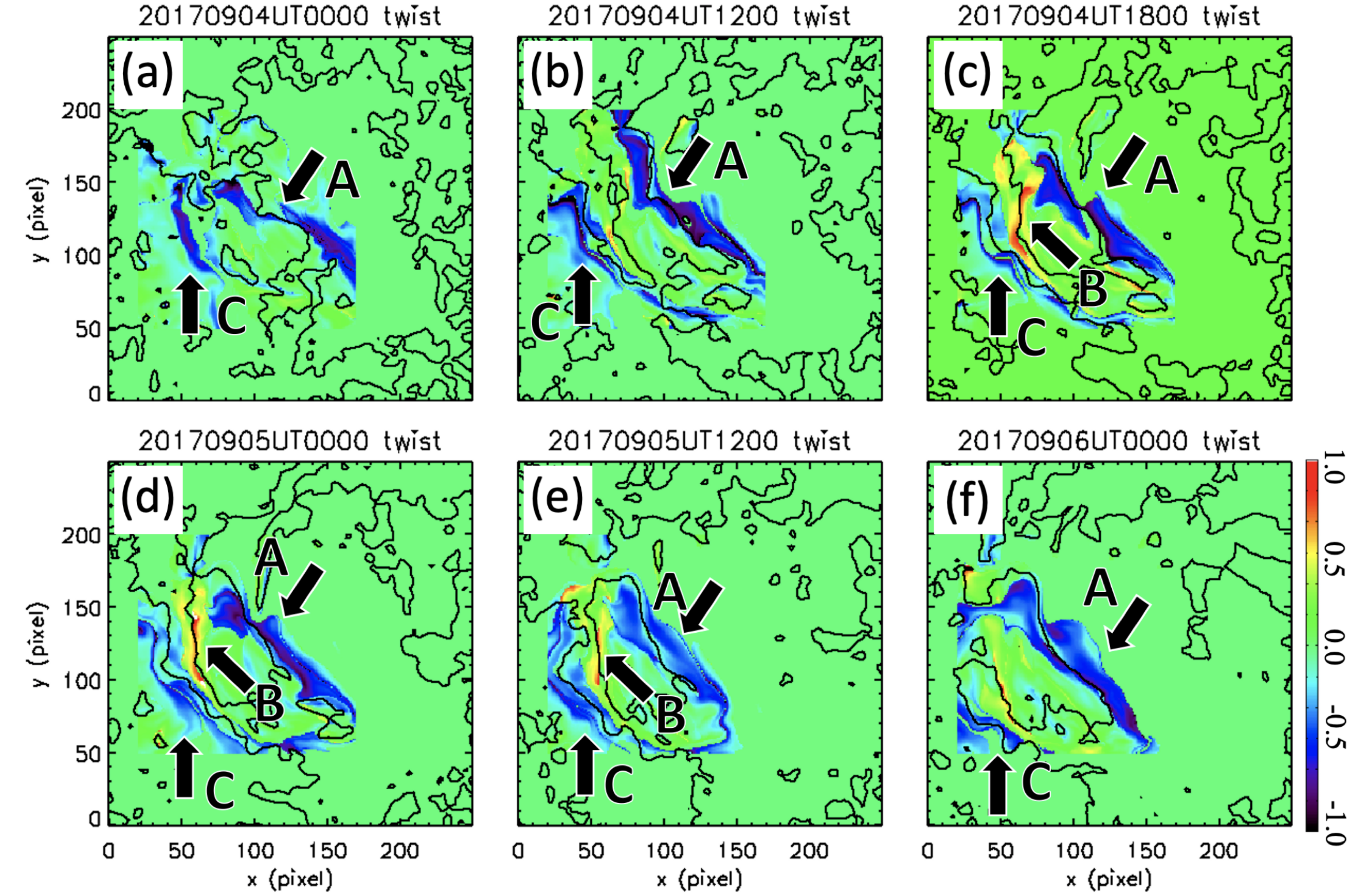}
\caption{Temporal evolution of the magnetic twist of field lines calculated from September 4th 00:00 UT to September 6th 00:00 UT. Magnetic twist is calculated from each field line and mapped on the photosphere. The color scale represents the amplitude of magnetic twist but the positive and negative colors correspond to left- and right-handed twist. The black lines correspond to the PILs.}
\label{fig3}
\end{center}
\end{figure*}

We employ the magnetic twist, introduced by \citet{Berger2006}, \citet{Inoue2011}, and \citet{Liu2016} in this study because it is a convenient value to measure the twist of the field lines in each magnetic field line quantitatively.
The magnetic twist is defined as 
\begin{eqnarray}
  T_{\mathrm{w}} = \int_{L} \frac{\mu_{0}\bm{J_{\parallel}}}{4\pi\bm{B}}dl = \int_{L}\frac{\bm{\nabla}\times\bm{B}\cdot\bm{B}}{4\pi B^2}~ dl.  
\end{eqnarray}
We calculated the magnetic twist for each field line obtained from the NLFFF.\\
~ Figure \ref{fig3} shows the temporal evolution of the magnetic twist mapped on the photosphere.
According to the figires \ref{fig2} and \ref{fig3}, the MFRs A and C are characterized by the negative twist, while the MFR B has positive twist.
The value of positive twist peaks at 00:00 UT on September 5th, after which it seems to decrease. 
This result indicates that MFR B are formed at the site between the PILs where the MFRs A and C reside.

\begin{figure*}[htb]
\begin{center}
\includegraphics[bb= 0 0 957.6 848.64, width=150mm]{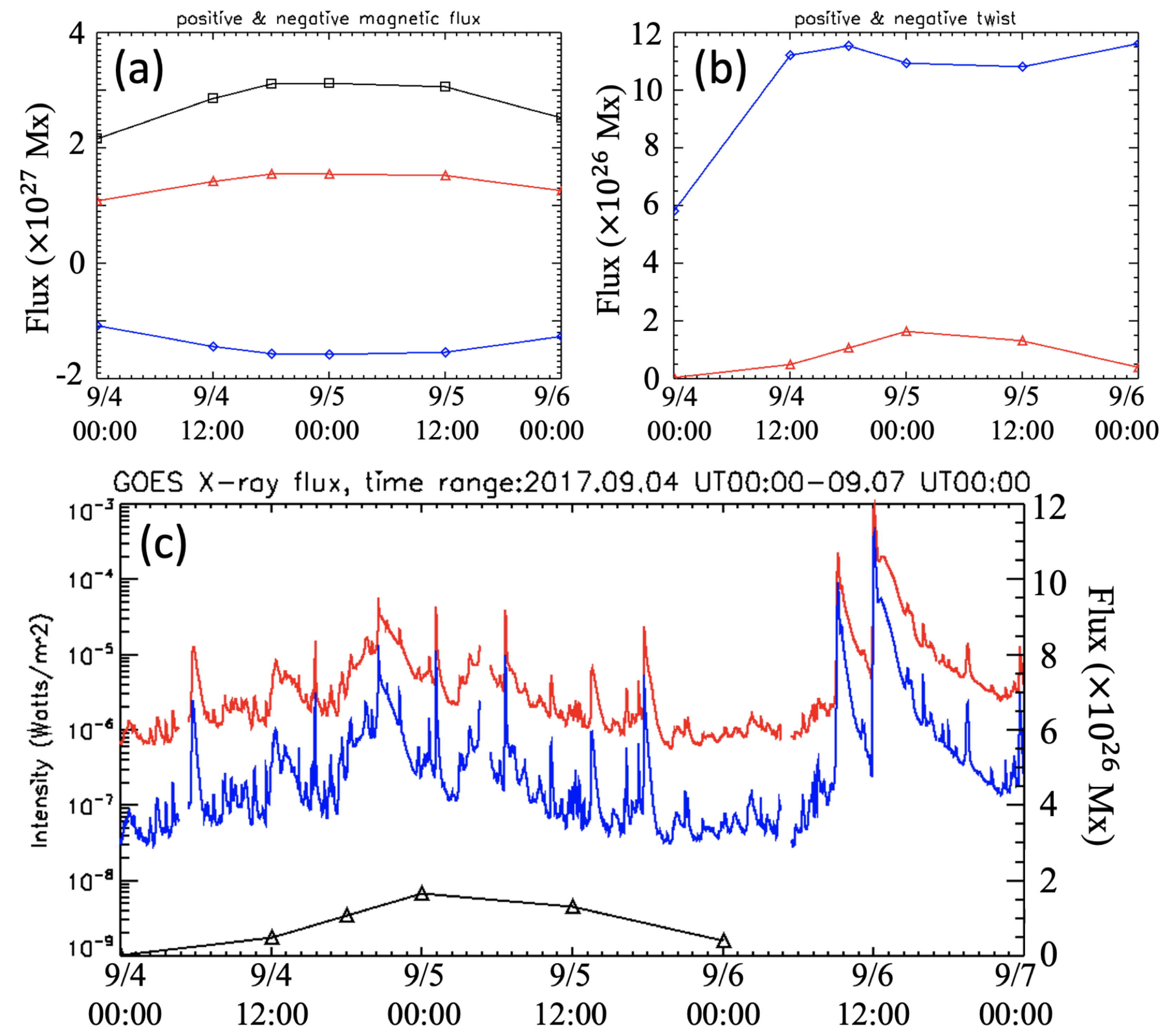}
\caption{(a) Temporal evolution of the magnetic flux in AR 12673. Red, blue, and black symbols represent positive, negative, and total unsigned magnetic flux respectively. (b) Temporal evolution of $\tau_{\pm}$ which are defined in equations (10) and (11). $\tau_{+}$ and $\tau_{-}$ are represented in red and blue, respectively. (c) Temporal evolution of $\tau_{+}$ is plotted over the evolution of the $GOES$ X-ray flux which is same as Figure \ref{fig1} (a).}
\label{fig4}
\end{center}
\end{figure*}

\begin{figure*}[htb]
\begin{center}
\includegraphics[bb= 0 0 817.93 759.84, width=150mm]{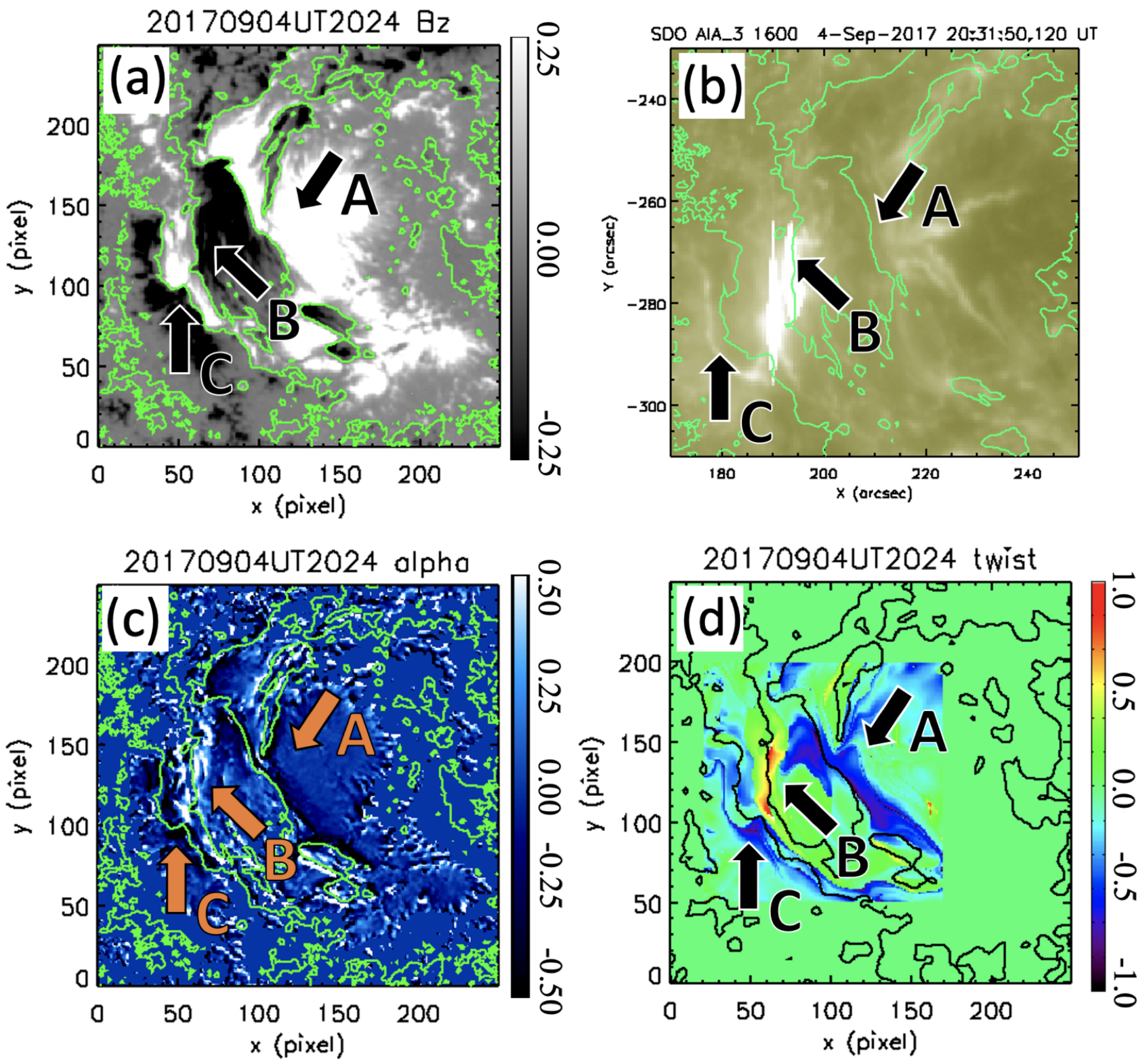}
\caption{(a) Photospheric magnetic field strendth at 20:24 UT on September 4th, green contour represents the PIL. (b) AIA 1600 $\mathrm{\AA}$ image at 20:31 UT on September 4th during the M5.5 flare. (c) Distribution of force-free alpha map calculated from the observed vector magnetic field at 20:24 UT on September 4th, green contour is same as (a). (d) Spatial distribution of the magnetic twist at 20:24 UT on September 4th, black contour represents the PIL.}
\label{fig5r}
\end{center}
\end{figure*}

In order to understand the evolution of the magnetic field more quantitatively, we plot the unsigned magnetic flux as follows, 
\begin{eqnarray}
  F_{+}(t) &=& \int_{B_{\mathrm{Z}}(t)>0}B_{\mathrm{Z}}(t) ~ dS, \\
  F_{-}(t) &=& \int_{B_{\mathrm{Z}}(t)<0}B_{\mathrm{Z}}(t) ~ dS, \\
  F_{\mathrm{tot}}(t) &=& F_{+}(t)+|F_{-}(t)|. 
\end{eqnarray}
\\
~ Figure \ref{fig4} (a) shows the result of the temporal evolution of the flux, $F_{+}$, $F_{-}$, and $F_{\mathrm{tot}}$ respectively. 
According to the figure \ref{fig4} (a), both positive and negative magnetic flux increase from September 4th 00:00 UT to September 5th 00:00 UT and decrease from September 5th 12:00 UT to September 6th 00:00 UT.
These results are consistent with the temporal evolution of the net flux reported by \citet{Vemareddy2019}. \\
~ Next, we calculated the twist flux, $\tau_{\pm}$, defined as followings, 
\begin{eqnarray}
  \tau_{+}(t) &=& \int_{T_{\mathrm{w}}(t)>0.35}|B_{\mathrm{Z}}(t)|\cdot T_{\mathrm{w}}(t) ~ dS,\\
  \tau_{-}(t) &=& \int_{T_{\mathrm{w}}(t)<-0.35}|B_{\mathrm{Z}}(t)|\cdot T_{\mathrm{w}}(t) ~ dS.
\end{eqnarray}
This value is a proxy of strength of a MFR.
Figure \ref{fig4} (b) shows the temporal evolution of $\tau_{+}$ and $\tau_{-}$. 
The evolution of $\tau_{-}$ clearly shows a rapid evolution within a day and keeps a high value until September 6th.
However, although $\tau_{+}$ also grows within a day, it takes the peak at September 5th 00:00 UT, and decreases after that.
We also compare the temporal evolution of $\tau_{+}$ with the $GOES$ soft X-ray profiles in figure \ref{fig4} (c).
The evolution of $\tau_{+}$ well captures the evolution of the $GOES$ flux as shown in a period from September 4th to 6th.
In addition, we show the AIA 1600 $\mathrm{\AA}$ image, the spatial distributions of force-free alpha, and the magnetic twist around 20:30 UT on September 4th, when $\tau_{+}$ increases, in figures \ref{fig5r} (b), (c), and (d), respectively. Here we note that force-free alpha ($\alpha$) was calculated with the photospheric vector magnetic field obtained with the HMI according to $\alpha=(\bm{\nabla}\times\bm{B})_{\mathrm{z}}/B_{\mathrm{z}}$.
In the figures, we can clearly see that the sign of force-free alpha and the magnetic twist show the similar distribution. Especially, around the MFR B, both force-free alpha and the magnetic twist have positive sign. Additionally, the enhancement of AIA 1600 $\mathrm{\AA}$ was observed at the site of the MFR B. These results, from both modeling and observations, suggest that the evolution of the MFR B having the positive twist is closely related to the occurrence of the successive M- and C-class flares.


\subsubsection{Temporal evolution of the AIA 1600$\mathrm{\AA}$ observations along each PIL} \label{subsec:teoa}
In this section, we analyze the AIA 1600 $\mathrm{\AA}$ image to identify the location of the flares.
\begin{figure*}[h]
\begin{center}
\includegraphics[bb= 0 0 629.76 848.64, width=150mm]{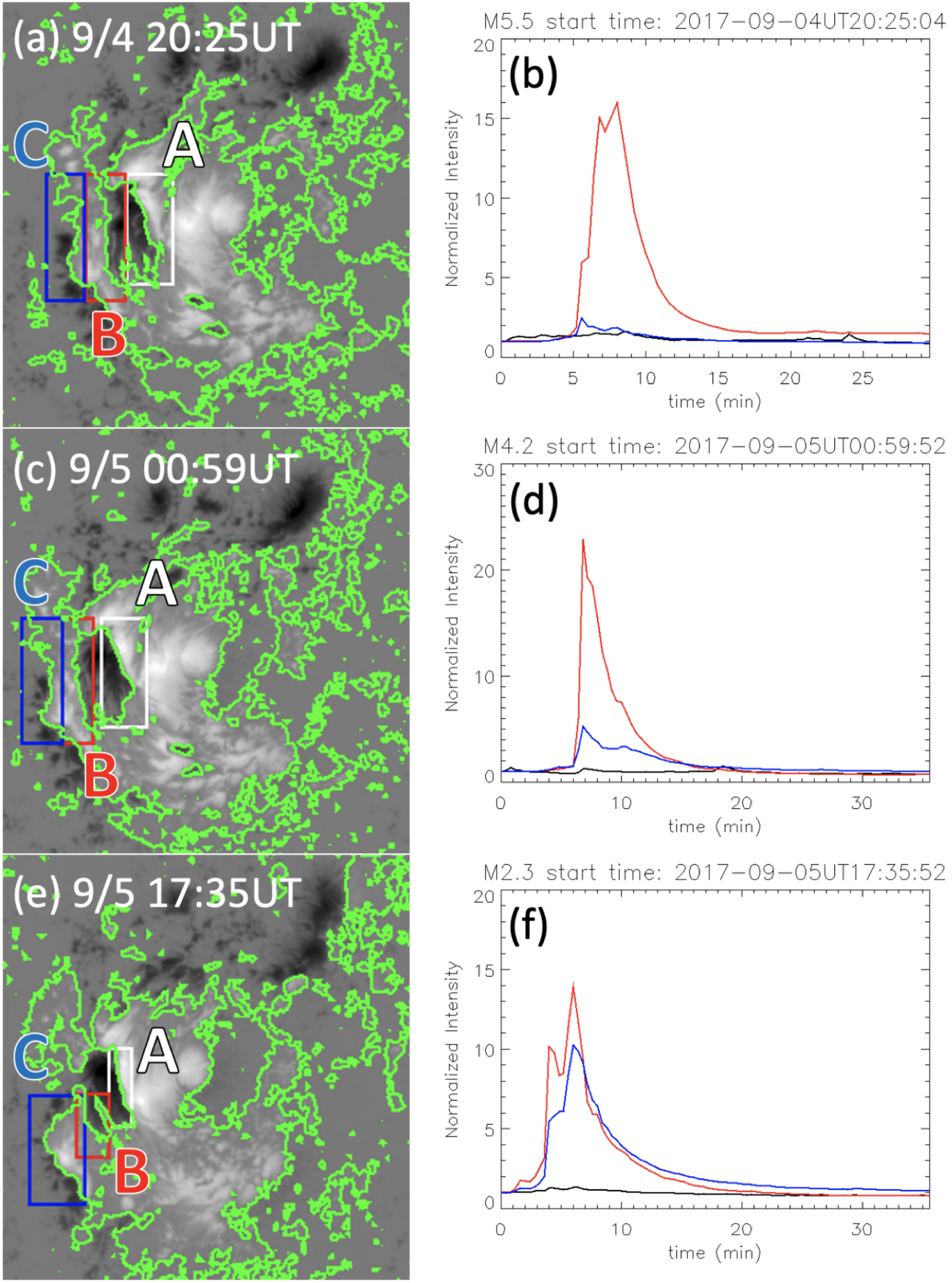}
\caption{Left three panels (a), (c), and (e) show the HMI magnetic field strength in grayscale, and PILs with green lines. These are observed at 04:24 UT September 4th, 01:00 UT, and 17:36 UT September 5th, respectively. Right three panels (b), (d), and (f) show the temporal evolution of intensity of the AIA 1600 $\mathrm{\AA}$, which is integrated in each box normalized by the intensity at the start time during the solar flares. Each red, blue, and black line is calculated in each integration area in red , blue, and white box respectively, shown in left panels.}
\label{fig5}
\end{center}
\end{figure*}
From the GOES plot, we see that 5 flares greater than M2.0-class were observed from September 4th 18:00 UT to September 6th 00:00 UT. 
We show the results from three of them: M5.5 flare observed at 20:30 UT September 4th, M4.2 flare observed at 01:05 UT September 5th, and M2.3 flare observed at 17:40 UT September 5th. 
We divided the flaring region into three areas along three PILs which are surrounded by squares as shown in figures \ref{fig5} (a), (c), and (e). 
We measure the intensity of AIA 1600 $\mathrm{\AA}$ at each box.
The results are shown in figures \ref{fig5} (b), (d), and (f). 
The first two M-class flares obviously occur on PIL B located in the center, i.e., the MFR B having positive twist would cause these flares.
Therefore, we can suggest that the M5.5 flare is mainly driven by the MFR having positive twist.
Regarding the last M-flare, the intensity is not only observed at the same PIL but also at the PIL C located in the east.
Interestingly, the AIA intensity profile at PIL A located in west was quiet during this period.
The MFR residing there produced the successive X-class flares later.
This result suggests that the MFR can accumulate the free energy during this period without being interrupted. 

\subsection{Magnetic field structure which produced the M5.5 flare}
As shown in figure \ref{fig4} (c), the temporal evolutions of $GOES$ X-ray flux and the magnetic flux of the MFR dominated by the positive twist, $\tau_{+}$, seem to have a good correlation with each other.
This result suggests the formation of the MFR B, having positive twist, closely relates to the occurrence of M- and C-class flares. 
However, the exact trigger mechanism of these flares is not yet clear.
Therefore, we investigate the initiation, in particular, focusing on the M5.5 flare which is the largest flare among multiple M-flares observed from September 4th to 6th, and also produced a geo-effective CME.

\subsubsection{Comparison between NLFFF and the AIA 1600 $\mathrm{\AA}$ image before the M5.5 flare}
\begin{figure*}[htb]
  \begin{center}
    \includegraphics[bb= 0 0 1169.76 566.88, width=150mm]{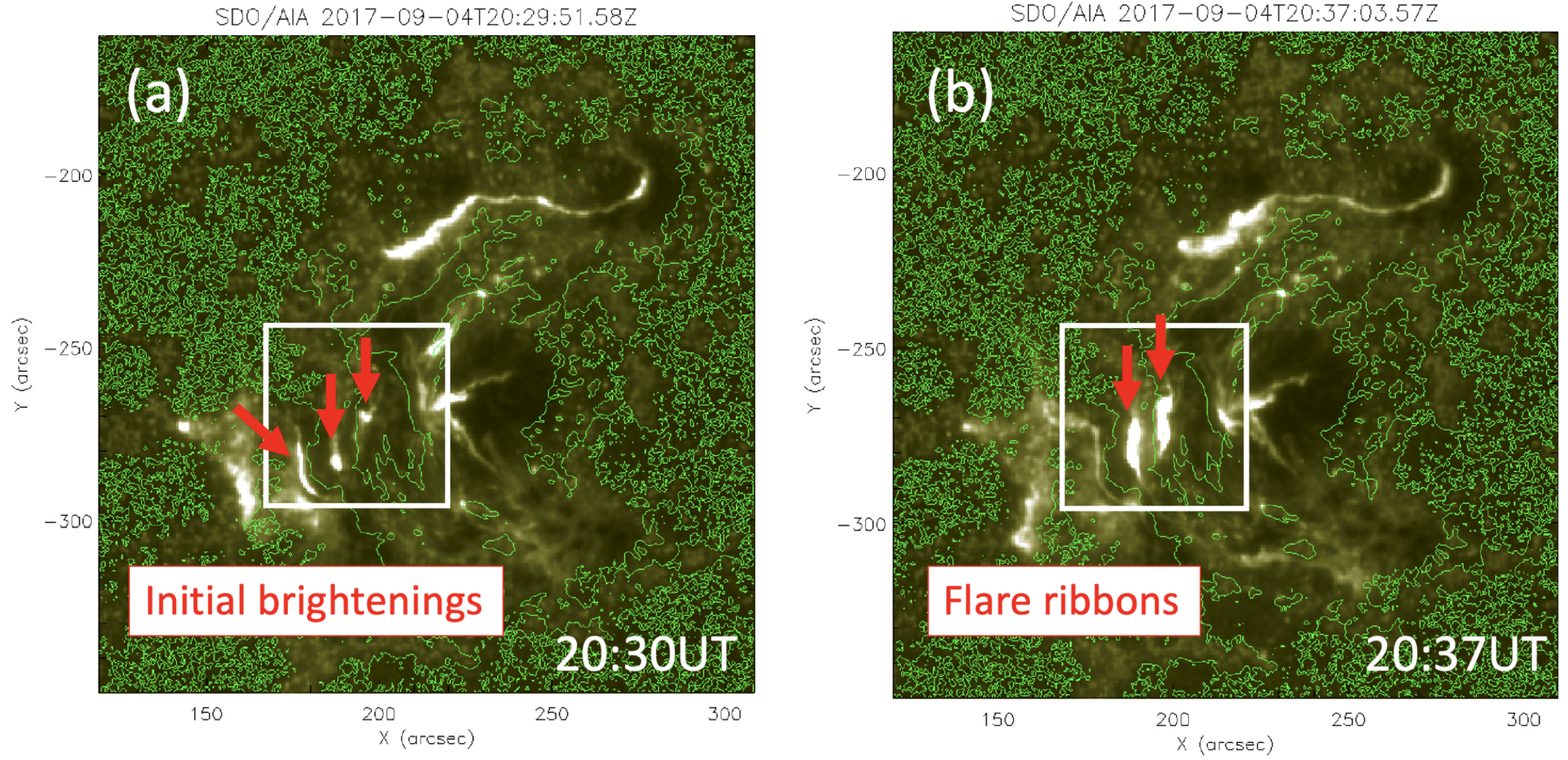}
    \caption{(a) AIA 1600 $\mathrm{\AA}$ image when the M5.5 flare started at September 4th 20:30 UT where the green lines correspond to the PILs. (b) Flare ribbons of M5.5 flare observed at September 4th 20:37 UT in AIA $1600$ $\mathrm{\AA}$ image.}
    \label{fig6}
  \end{center}
\end{figure*}

\begin{figure*}[htb]
  \begin{center}
    \includegraphics[bb= 0 0 974.88 660.96, width=150mm]{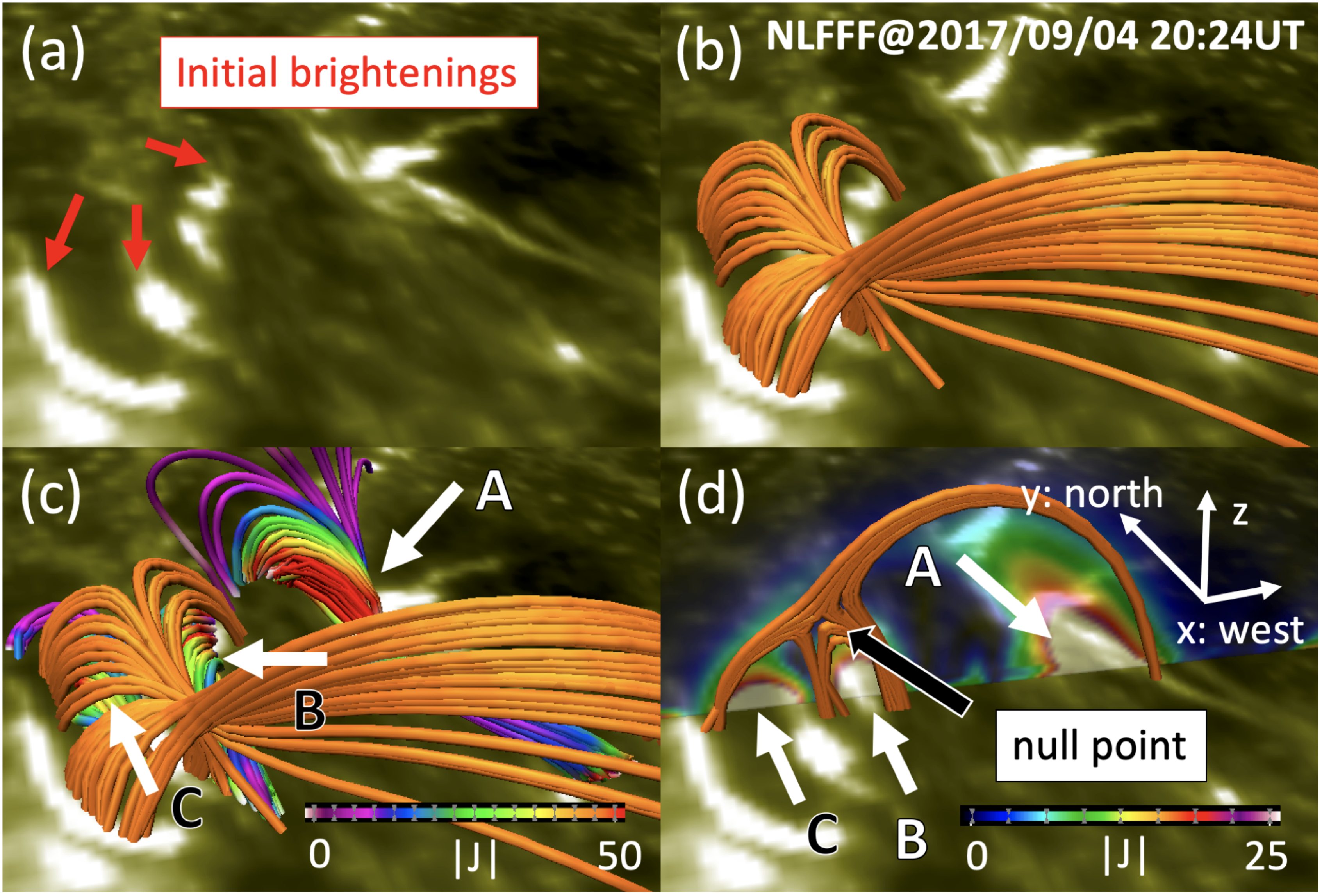}
    \caption{(a) AIA 1600 $\mathrm{\AA}$ image at the beginning of the M5.5 flare. Red arrows indicate the brightening associated with initiation of the M5.5 flare. (b) The field lines obtained from the NLFFF extrapolation using the HMI data at 20:24 UT on September 4th, are traced from the three brightening regions indicated by arrows shown in (a). (c) Three MFRs plotted together with the field lines shown in (b). (d) The field lines surrounding the MFRs, which correspond to those shown in (b), and they are traced in $x-z$ plane where $y=0.5$. The distribution of electric current density in the same plane.}
    \label{fig7}
  \end{center}
\end{figure*}

Figures \ref{fig6} (a) and (b) show the evolution of the flare ribbons observed in the beginning phase of the M5.5 flare. 
The initial brightenings of the flare observed at 20:30 UT occur at the three locations indicated by red arrows in the white square shown in figure \ref{fig6} (a).
These are enhanced on the both sides of the eastern PIL and central PIL, respectively.
However, in the early phase of the flare (at 20:37 UT), the two flare ribbons grow on both side of the center PIL as shown in figure \ref{fig6} (b).
Figure \ref{fig7} (a) shows a close-up of the AIA 1600 $\mathrm{\AA}$ image at 20:30 UT in which the initial brightenings of the M5.5 flare are found. 
In figure \ref{fig7} (b), we traced the field lines from the region where the initial brightenings are enhanced, i.e., these field lines shown in orange are related to the onset of the M5.5 flare.
Figure \ref{fig7} (c) shows field lines that form the three MFRs in addition to the field lines shown in figure \ref{fig7} (b).
It is found that the MFRs are covered with the orange lines.
For a better understanding of the field configuration, we produced figure \ref{fig7} (d) in which the field lines are traced only in the $x-z$ plane at $y=0.5$, and on which the current density $|\bm{J}|$ is displayed with colors. 
The regions of enhanced current $|\bm{J}|$ correspond to the regions where the MFRs exist.
We also found that a magnetic null point exists above MFR B.

\subsubsection{Decay index distribution above the MFRs}
\begin{figure*}[htb]
  \begin{center}
    \includegraphics[bb= 0 0 456 632.64, width=100mm]{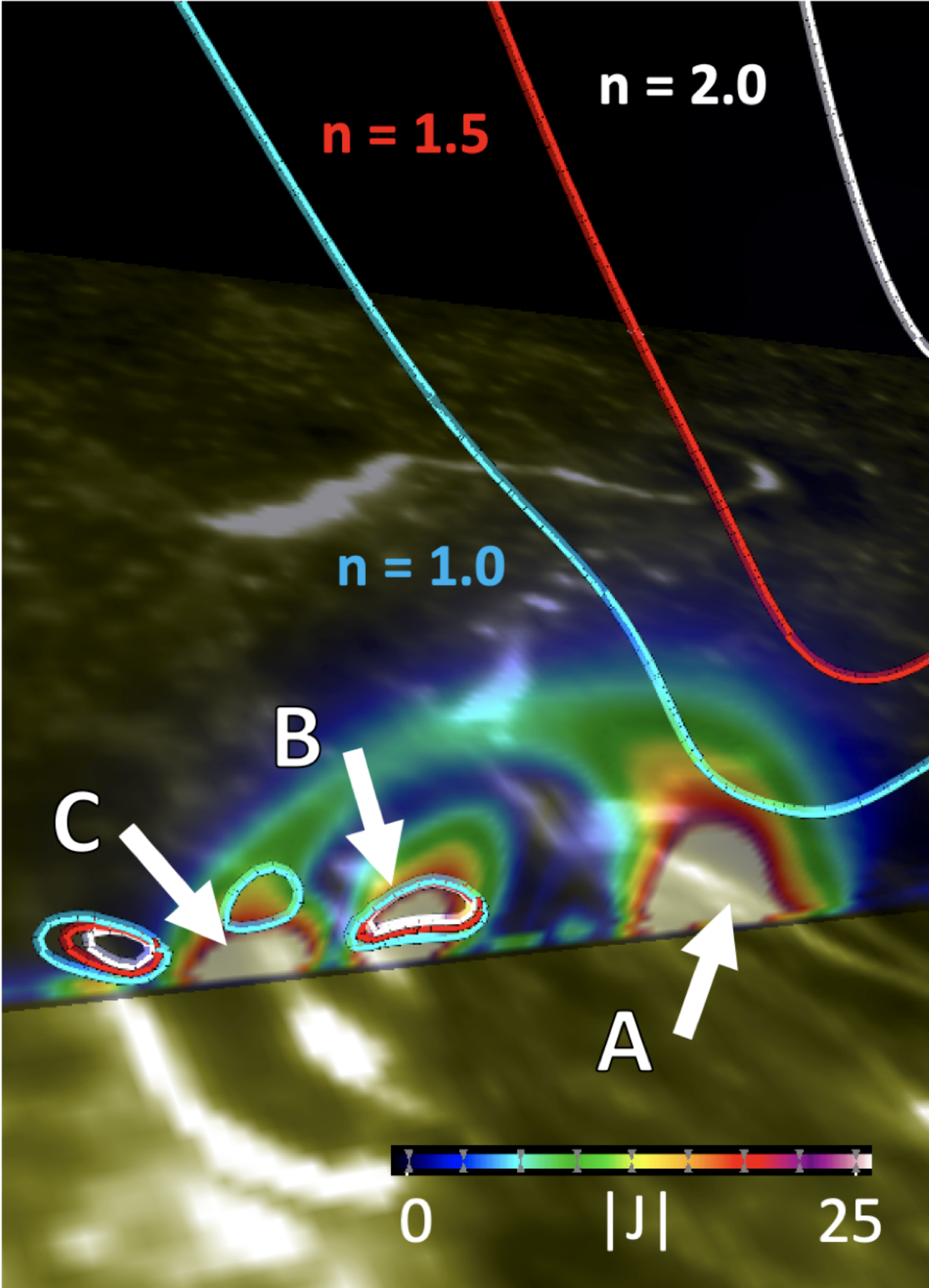}
    \caption{The contour of the decay index, $n=1.0$, $1.5$, and $2.0$ are shown in blue, red, and white, respectively on the $x-z$ plane at $y=0.5$, on which the distribution of electric current density is displayed. The electrical current distrubution is the same as in figure \ref{fig7}.}
    \label{fig8}
  \end{center}
\end{figure*}

We further plot the decay index $n$, which is a proxy of the torus instability \citep{Kliem2006}, in figure \ref{fig8}.
The decay index $n$ is defined as $n=-(z/|\bm{B_{\mathrm{ex}}}|)(\partial |\bm{B_{\mathrm{ex}}}|/\partial z)$ where $\bm{B_{\mathrm{ex}}}$ denotes the horizontal component of the external field.
Here we asume that the external field is the potential field.
According to the figure, since the MFRs A and C reside outside of the region that satisfies $n>1.5$, we suggest that these MFRs are stable to the torus instability.
However, MFR B is located in the region where the decay index $n$ is larger than 1.5.

\section{Discussion \& Summary} \label{sec:disc}
\begin{figure*}[htb]
  \begin{center}
    \includegraphics[bb= 0 0 1196.64 437.76, width=150mm]{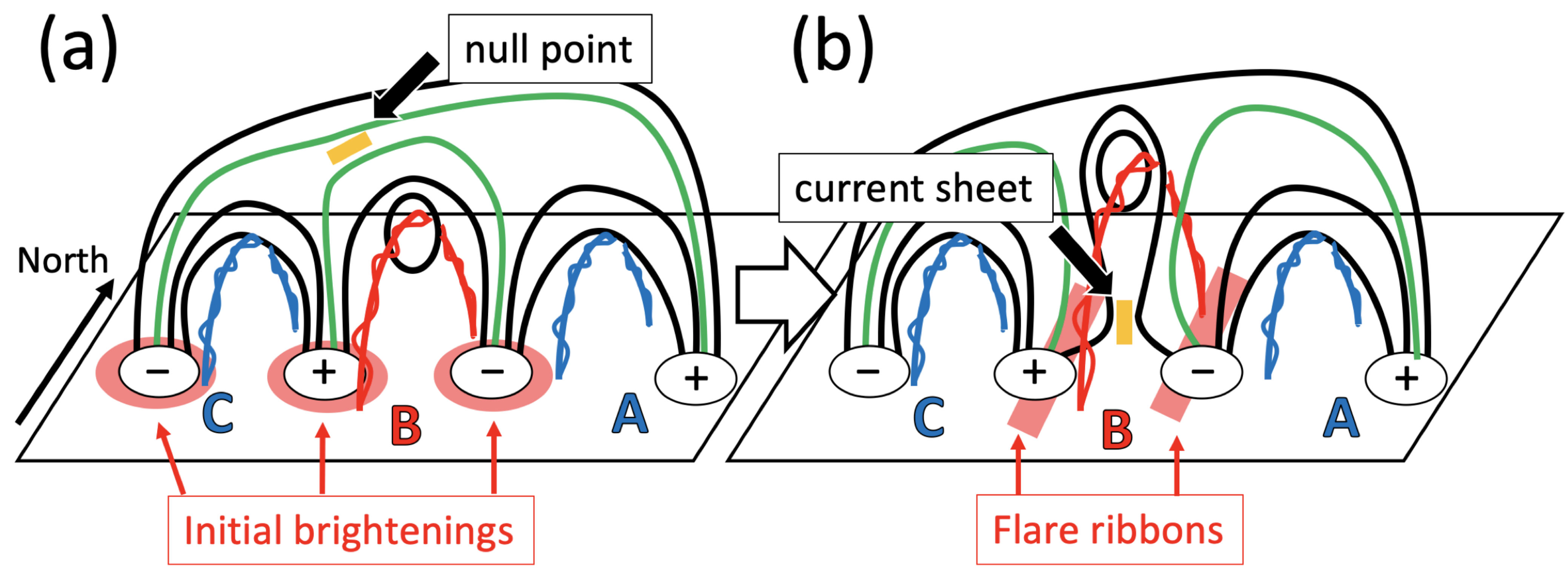}
    \caption{The schematic of the magnetic field structure before and after the M5.5 flare. Red and blue twisted lines represent the MFRs A, B, and C. The red and blue represent the sign of the helicity where the red and blue correspond to positive and negative, respectively. Green lines show the coronal magnetic field lines, which experience the reconnection at the magnetic null point in the early phase of M5.5 flare. Yellow line shows the current sheet. Red arrows in (a) indicate the location of initial brightenings, and those in (b) indicate the location of the flare ribbons.}
    \label{fig9}
  \end{center}
\end{figure*}

In this study, in order to clarify the evolution of the 3D magnetic field and flare activities observed in AR 12673, we analyzed the temporal evolution of the 3D magnetic field structure based on the time series of the NLFFFs from September 4th 00:00 UT to 6th 00:00 UT. 
We calculated the magnetic twist from each NLFFF, and consequently, found that three MFRs are formed above three PILs in parallel before the M5.5 flare, observed at 18:30 UT on September 4th.
Interestingly one MFR (MFR B) locating in the center has positive helicity while the other two (MFRs A and C) have negative helicity.
From our detailed analysis, the evolution of the magnetic flux of MFR B has a good correlation with the $GOES$ flux evolution while the M- and C-class flares were observed.
Therefore, we suggest that these flares were driven by the formation of the MFR B.
We further analyzed the 3D magnetic structure in vicinity of the three MFRs.
We found a magnetic null above MFR B at 20:24 UT on September 4th.
\citet{Vemareddy2019} also reported that a magnetic null was formed in a similar location at 18:30 UT on September 4th. 
Since the AIA $1600$ $\mathrm{\AA}$ image shows intensity enhancements in the beginning of the M5.5 flare where the footpoints of the magnetic field lines including magnetic null are anchored, we suggest that null point reconnection possibly helps to accelerate the MFR B. 
Furthermore, as shown in figure \ref{fig8}, we found that MFR B would be unstable to the torus instability.
\citet{Inoue2018a} showed that MFRs can be accelerated even in an isolated area where the decay index $n$ has high value to drive the instability.
Therefore, in this case, the torus instability possibly pushed the MFR B and enhanced the reconnection at the magnetic null.
Namely, the interaction between the torus instability and reconnection might be important driving the eruption. \\
~ Figure \ref{fig9} schematically describes a scenario for the M5.5 flare obtained from our study.
When the positive twisted MFR B is formed at the site between the MFRs A and C, it pushes the pre-existing magnetic field upward which drives the reconnection at the null point.
The null point reconnection is able to remove the magnetic field surrounding the MFR B, consequently, the MFR B can easily escape from the lower corona resulting in the eruption.
However, reconnected field lines would suppress the instabilityof neighboring MFRs A and C.
Therefore, only MFR B would escape from the lower corona. \\
~ However, we confirmed that the AIA $1600$ $\mathrm{\AA}$ brightening is relatively weak above the PIL located on the west side of the AR, compared to those located in the middle and east side. 
The MFR A located above the PIL in west causes successive X-class flares on September 6th \citep{Hou2018,Inoue2018b,Mitra2018}.
Therefore, we suggest that MFR A can accumulate the enough free magnetic energy to produce the X-flares as a result of photospheric motions without being disturbed.
In addition, at the west-side PIL the intruding motion of the negative peninsula toward the neighboring positive was observed before the X2.2 flare \citep{Bamba2020}.
Taking into account our results, this intrusion would be important to break the equilibrium of the MFR A, resulting in triggering the X-flares. \\
~ According to our study, we suggest that not only the properties of the MFRs but also the 3D magnetic topology overlying the MFR is important to understand the onset of solar flares.
Therefore, we consider that this information should be taken into account when considering future flare prediction schemes. \\

\acknowledgements
We thank the anonymous referee for helping us to improve and polish this paper. 
We are grateful to Dr. Yumi Bamba for providing AIA images processed into CEA projection. 
We thank Professor Kanya Kusano for useful discussions. 
We are grateful to Dr. Magnus Woods for checking this manuscript. 
$SDO$ is a mission of NASA’s Living With a Star Program.
This work was supported by MEXT/JSPS KAKENHI Grant Number JP15H05814, Project for Solar-Terrestrial Environment Prediction (PSTEP) and the computational joint research program of the Institute for Space-Earth Environmental Research (ISEE), Nagoya University. 
Visualization of magnetic field lines are produced by VAPOR (\url{www.vapor.ucar.edu}), a product of the Computational Information Systems Laboratory at the National Center for Atmospheric Research.

\bibliography{yamasaki2020_arXiv}

\begin{thebibliography}{}
\expandafter\ifx\csname natexlab\endcsname\relax\def\natexlab#1{#1}\fi
\providecommand{\url}[1]{\href{#1}{#1}}
\providecommand{\dodoi}[1]{doi:~\href{http://doi.org/#1}{\nolinkurl{#1}}}
\providecommand{\doeprint}[1]{\href{http://ascl.net/#1}{\nolinkurl{http://ascl.net/#1}}}
\providecommand{\doarXiv}[1]{\href{https://arxiv.org/abs/#1}{\nolinkurl{https://arxiv.org/abs/#1}}}

\bibitem[{{Bamba} {et~al.}(2020){Bamba}, {Inoue}, \& {Imada}}]{Bamba2020}
{Bamba}, Y., {Inoue}, S., \& {Imada}, S. 2020, \apj, 894, 29,
  \dodoi{10.3847/1538-4357/ab85ca}

\bibitem[{{Berger} \& {Prior}(2006)}]{Berger2006}
{Berger}, M.~A., \& {Prior}, C. 2006, Journal of Physics A Mathematical
  General, 39, 8321, \dodoi{10.1088/0305-4470/39/26/005}

\bibitem[{Bobra {et~al.}(2014)Bobra, Sun, Hoeksema, Turmon, Liu, Hayashi,
  Barnes, \& Leka}]{Bobra2014}
Bobra, M.~G., Sun, X., Hoeksema, J.~T., {et~al.} 2014, Solar Physics, 289,
  3549, \dodoi{10.1007/s11207-014-0529-3}

\bibitem[{Chen(2011)}]{Chen2011}
Chen, P.~F. 2011, Living Reviews in Solar Physics, 8, 1,
  \dodoi{10.12942/lrsp-2011-1}

\bibitem[{{Cheng} {et~al.}(2017){Cheng}, {Guo}, \& {Ding}}]{Cheng2017}
{Cheng}, X., {Guo}, Y., \& {Ding}, M. 2017, Science China Earth Sciences, 60,
  1383, \dodoi{10.1007/s11430-017-9074-6}

\bibitem[{{Chertok} {et~al.}(2018){Chertok}, {Belov}, \&
  {Abunin}}]{Chertok2018}
{Chertok}, I.~M., {Belov}, A.~V., \& {Abunin}, A.~A. 2018, Space Weather, 16,
  1549, \dodoi{10.1029/2018SW001899}

\bibitem[{{Dedner} {et~al.}(2002){Dedner}, {Kemm}, {Kr{\"o}ner}, {Munz},
  {Schnitzer}, \& {Wesenberg}}]{Dedner2002}
{Dedner}, A., {Kemm}, F., {Kr{\"o}ner}, D., {et~al.} 2002, Journal of
  Computational Physics, 175, 645, \dodoi{10.1006/jcph.2001.6961}

\bibitem[{{Filippov} {et~al.}(2015){Filippov}, {Martsenyuk}, {Srivastava}, \&
  {Uddin}}]{Filippov2015}
{Filippov}, B., {Martsenyuk}, O., {Srivastava}, A.~K., \& {Uddin}, W. 2015,
  Journal of Astrophysics and Astronomy, 36, 157,
  \dodoi{10.1007/s12036-015-9321-5}

\bibitem[{Guo {et~al.}(2017)Guo, Cheng, \& Ding}]{Guo2017}
Guo, Y., Cheng, X., \& Ding, M. 2017, Science China Earth Sciences, 60, 1408,
  \dodoi{10.1007/s11430-017-9081-x}

\bibitem[{Hoeksema {et~al.}(2014)Hoeksema, Liu, Hayashi, Sun, Schou, Couvidat,
  Norton, Bobra, Centeno, Leka, Barnes, \& Turmon}]{Hoeksema2014}
Hoeksema, J.~T., Liu, Y., Hayashi, K., {et~al.} 2014, Solar Physics, 289, 3483,
  \dodoi{10.1007/s11207-014-0516-8}

\bibitem[{{Hou} {et~al.}(2018){Hou}, {Zhang}, {Li}, {Yang}, \& {Li}}]{Hou2018}
{Hou}, Y.~J., {Zhang}, J., {Li}, T., {Yang}, S.~H., \& {Li}, X.~H. 2018, \aap,
  619, A100, \dodoi{10.1051/0004-6361/201732530}

\bibitem[{{Inoue}(2016)}]{Inoue2016}
{Inoue}, S. 2016, Progress in Earth and Planetary Science, 3, 19,
  \dodoi{10.1186/s40645-016-0084-7}

\bibitem[{{Inoue} {et~al.}(2018a){Inoue}, {Kusano}, {B{\"u}chner}, \&
  {Sk{\'a}la}}]{Inoue2018a}
{Inoue}, S., {Kusano}, K., {B{\"u}chner}, J., \& {Sk{\'a}la}, J. 2018a, Nature
  Communications, 9, 174, \dodoi{10.1038/s41467-017-02616-8}

\bibitem[{{Inoue} {et~al.}(2011){Inoue}, {Kusano}, {Magara}, {Shiota}, \&
  {Yamamoto}}]{Inoue2011}
{Inoue}, S., {Kusano}, K., {Magara}, T., {Shiota}, D., \& {Yamamoto}, T.~T.
  2011, \apj, 738, 161, \dodoi{10.1088/0004-637X/738/2/161}

\bibitem[{{Inoue} {et~al.}(2014){Inoue}, {Magara}, {Pandey}, {Shiota},
  {Kusano}, {Choe}, \& {Kim}}]{Inoue2014a}
{Inoue}, S., {Magara}, T., {Pandey}, V.~S., {et~al.} 2014, \apj, 780, 101,
  \dodoi{10.1088/0004-637X/780/1/101}

\bibitem[{{Inoue} {et~al.}(2018b){Inoue}, {Shiota}, {Bamba}, \&
  {Park}}]{Inoue2018b}
{Inoue}, S., {Shiota}, D., {Bamba}, Y., \& {Park}, S.-H. 2018b, \apj, 867, 83,
  \dodoi{10.3847/1538-4357/aae079}

\bibitem[{{Kliem} \& {T{\"o}r{\"o}k}(2006)}]{Kliem2006}
{Kliem}, B., \& {T{\"o}r{\"o}k}, T. 2006, \prl, 96, 255002,
  \dodoi{10.1103/PhysRevLett.96.255002}

\bibitem[{Lemen {et~al.}(2012)Lemen, Title, Akin, Boerner, Chou, Drake, Duncan,
  Edwards, Friedlaender, Heyman, Hurlburt, Katz, Kushner, Levay, Lindgren,
  Mathur, McFeaters, Mitchell, Rehse, Schrijver, Springer, Stern, Tarbell,
  Wuelser, Wolfson, Yanari, Bookbinder, Cheimets, Caldwell, Deluca, Gates,
  Golub, Park, Podgorski, Bush, Scherrer, Gummin, Smith, Auker, Jerram, Pool,
  Soufli, Windt, Beardsley, Clapp, Lang, \& Waltham}]{Lemen2012}
Lemen, J.~R., Title, A.~M., Akin, D.~J., {et~al.} 2012, Solar Physics, 275, 17,
  \dodoi{10.1007/s11207-011-9776-8}

\bibitem[{{Liu} {et~al.}(2019){Liu}, {Cheng}, {Wang}, \& {Zhou}}]{Liu2019}
{Liu}, L., {Cheng}, X., {Wang}, Y., \& {Zhou}, Z. 2019, \apj, 884, 45,
  \dodoi{10.3847/1538-4357/ab3c6c}

\bibitem[{{Liu} {et~al.}(2016){Liu}, {Kliem}, {Titov}, {Chen}, {Wang}, {Wang},
  {Liu}, {Xu}, \& {Wiegelmann}}]{Liu2016}
{Liu}, R., {Kliem}, B., {Titov}, V.~S., {et~al.} 2016, \apj, 818, 148,
  \dodoi{10.3847/0004-637X/818/2/148}

\bibitem[{{Mitra} {et~al.}(2018){Mitra}, {Joshi}, {Prasad}, {Veronig}, \&
  {Bhattacharyya}}]{Mitra2018}
{Mitra}, P.~K., {Joshi}, B., {Prasad}, A., {Veronig}, A.~M., \&
  {Bhattacharyya}, R. 2018, \apj, 869, 69, \dodoi{10.3847/1538-4357/aaed26}

\bibitem[{{Okamoto} {et~al.}(2008){Okamoto}, {Tsuneta}, {Lites}, {Kubo},
  {Yokoyama}, {Berger}, {Ichimoto}, {Katsukawa}, {Nagata}, {Shibata},
  {Shimizu}, {Shine}, {Suematsu}, {Tarbell}, \& {Title}}]{Okamoto2008}
{Okamoto}, T.~J., {Tsuneta}, S., {Lites}, B.~W., {et~al.} 2008, \apjl, 673,
  L215, \dodoi{10.1086/528792}

\bibitem[{Pesnell {et~al.}(2012)Pesnell, Thompson, \& Chamberlin}]{Pesnell2012}
Pesnell, W.~D., Thompson, B.~J., \& Chamberlin, P.~C. 2012, Solar Physics, 275,
  3, \dodoi{10.1007/s11207-011-9841-3}

\bibitem[{{Priest} \& {Forbes}(2002)}]{Priest2002}
{Priest}, E.~R., \& {Forbes}, T.~G. 2002, \aapr, 10, 313,
  \dodoi{10.1007/s001590100013}

\bibitem[{Sakurai(1982)}]{Sakurai1982}
Sakurai, T. 1982, Solar Physics, 76, 301, \dodoi{10.1007/BF00170988}

\bibitem[{Scherrer {et~al.}(2012)Scherrer, Schou, Bush, Kosovichev, Bogart,
  Hoeksema, Liu, Duvall, Zhao, Title, Schrijver, Tarbell, \&
  Tomczyk}]{Scherrer2012}
Scherrer, P.~H., Schou, J., Bush, R.~I., {et~al.} 2012, Solar Physics, 275,
  207, \dodoi{10.1007/s11207-011-9834-2}

\bibitem[{Schmieder {et~al.}(2015)Schmieder, Aulanier, \&
  Vr{\v{s}}nak}]{Schmieder2015}
Schmieder, B., Aulanier, G., \& Vr{\v{s}}nak, B. 2015, Solar Physics, 290,
  3457, \dodoi{10.1007/s11207-015-0712-1}

\bibitem[{{Shen} {et~al.}(2018){Shen}, {Xu}, {Wang}, {Chi}, \&
  {Luo}}]{Shen2018}
{Shen}, C., {Xu}, M., {Wang}, Y., {Chi}, Y., \& {Luo}, B. 2018, \apj, 861, 28,
  \dodoi{10.3847/1538-4357/aac204}

\bibitem[{Shibata \& Magara(2011)}]{Shibata2011}
Shibata, K., \& Magara, T. 2011, Living Reviews in Solar Physics, 8, 6,
  \dodoi{10.12942/lrsp-2011-6}

\bibitem[{{Soni} {et~al.}(2020){Soni}, {Gupta}, \& {Verma}}]{Soni2020}
{Soni}, S.~L., {Gupta}, R.~S., \& {Verma}, P.~L. 2020, Research in Astronomy
  and Astrophysics, 20, 023, \dodoi{10.1088/1674-4527/20/2/23}

\bibitem[{{van Ballegooijen} \& {Martens}(1989)}]{vanBallegooijen1989}
{van Ballegooijen}, A.~A., \& {Martens}, P.~C.~H. 1989, \apj, 343, 971,
  \dodoi{10.1086/167766}

\bibitem[{{Vemareddy}(2019)}]{Vemareddy2019}
{Vemareddy}, P. 2019, \apj, 872, 182, \dodoi{10.3847/1538-4357/ab0200}

\bibitem[{Wiegelmann {et~al.}(2006)Wiegelmann, Inhester, \&
  Sakurai}]{Wiegelmann2006}
Wiegelmann, T., Inhester, B., \& Sakurai, T. 2006, Solar Physics, 233, 215,
  \dodoi{10.1007/s11207-006-2092-z}

\bibitem[{Wiegelmann \& Sakurai(2012)}]{Wiegelmann2012}
Wiegelmann, T., \& Sakurai, T. 2012, Living Reviews in Solar Physics, 9, 5,
  \dodoi{10.12942/lrsp-2012-5}

\bibitem[{{Yan} {et~al.}(2018){Yan}, {Wang}, {Pan}, {Kong}, {Xue}, {Yang},
  {Li}, \& {Feng}}]{Yan2018}
{Yan}, X.~L., {Wang}, J.~C., {Pan}, G.~M., {et~al.} 2018, \apj, 856, 79,
  \dodoi{10.3847/1538-4357/aab153}

\bibitem[{{Yang} {et~al.}(2017){Yang}, {Zhang}, {Zhu}, \& {Song}}]{Yang2017}
{Yang}, S., {Zhang}, J., {Zhu}, X., \& {Song}, Q. 2017, \apjl, 849, L21,
  \dodoi{10.3847/2041-8213/aa9476}

\end{thebibliography}
\bibliographystyle{aasjournal}

\end{document}